\documentclass[11pt]{article}

\usepackage{graphicx}
\usepackage{acronym}
\usepackage{amsmath} 
\usepackage{multirow}
\usepackage{cite}
\usepackage{algorithmic}
\usepackage{algorithm}

\usepackage{flushend}

\newcommand{\eg}{\emph{e.g.}}

\newcommand{\ea}{\emph{et al.}}

\acrodef{OPEX}[OPEX]{Operating Expenses}
\acrodef{UE}[UE]{User Equipment}
\acrodef{BS}[BS]{Base Station}
\acrodef{DTX}[DTX]{Discontinuous Transmission}
\acrodef{PAPR}[PAPR]{Peak-to-Average Power Ratio }
\acrodef{SC-FDMA}[SC-FDMA]{Single-carrier FDMA}
\acrodef{FDMA}[FDMA]{Frequency Division Multiple Access}
\acrodef{TDMA}[TDMA]{Time Division Multiple Access}
\acrodef{CDMA}[CDMA]{Code Division Multiple Access}
\acrodef{OFDMA}[OFDMA]{Orthogonal Frequency Division Multiple Access}
\acrodef{ICT}[ICT]{Information and Communication Technologies}
\acrodef{QoS}[QoS]{Quality of Service}
\acrodef{PA}[PA]{Power Amplifier}
\acrodef{RS}[RS]{Resource Sharing}
\acrodef{PC}[PC]{Power Control}
\acrodef{SOTA}[SotA]{State-Of-The-Art}
\acrodef{EE}[EE]{Energy Efficiency}
\acrodef{SINR}[SINR]{Signal-to-Interference-and-Noise-Ratio}
\acrodef{LTE}[LTE]{Long Term Evolution}
\acrodef{EARTH}[EARTH]{Energy Aware Radio and neTwork tecHnologies}
\acrodef{MIMO}[MIMO]{Multiple-Input Multiple-Output}
\acrodef{SISO}[SISO]{Single-Input Single-Output}
\acrodef{SIMO}[SIMO]{Single-Input Multiple-Output}
\acrodef{RE}[RE]{Resource Element}
\acrodef{RB}[RB]{Resource Block}
\acrodef{SNR}[SNR]{Signal-to-Noise-Ratio}
\acrodef{ACLR}[ACLR]{Adjacent Carrier Leakage Ratio}
\acrodef{PRAIS}[PRAIS]{Power and Resource Allocation Including Sleep}
\acrodef{MT}[MT]{Mobile Terminal}
\acrodef{BA}[BA]{Bandwidth Adaptation}
\acrodef{MCS}[MCS]{Modulation and Coding Scheme}
\acrodef{AA}[AA]{Antenna Adaptation}
\acrodef{RRM}[RRM]{Radio Resource Management}
\acrodef{WSN}[WSN]{Wireless Sensor Network}
\acrodef{CSI}[CSI]{Channel State Information}
\acrodef{RRH}[RRH]{Remote Radio Head}
\acrodef{RCG}[RCG]{Rate Craving Greedy}
\acrodef{RF}[RF]{Radio Frequency}
\acrodef{3GPP}[3GPP]{3rd Generation Partnership Project}
\acrodef{RAPS}[RAPS]{Resource allocation using Antenna adaptation, Power control and Sleep modes}
\acrodef{MS}[MS]{Mains Supply}
\acrodef{BB}[BB]{Baseband}

\begin{document}

\title{Distributed DTX Alignment with Memory}


\author{{Hauke~Holtkamp, Guido Dietl}\vspace{3mm}\\
DOCOMO Euro-Labs\\
D-80687 Munich, Germany \\Email:
\{holtkamp, auer\}@docomolab-euro.com
\and Harald~Haas\vspace{2mm}\\
Institute for Digital Communications\\
Joint Research Institute for Signal and Image Processing\\
 The University of Edinburgh,
EH9 3JL, Edinburgh, UK\\ E-mail: h.haas@ed.ac.uk}

\addtolength{\textfloatsep}{-5mm}

\maketitle

\begin{abstract}
This paper addresses the assignment of transmission and sleep time slots between interfering transmitters with the objective of minimal power consumption. In particular, we address the constructive alignment of \ac{DTX} time slots under link rate constraints. Due to the complexity of the combinatorial optimization problem at hand, we resort to heuristic assignment strategies. We derive four time slot alignment solutions (sequential alignment, random alignment, p-persistent ranking and \ac{DTX} alignment with memory) and identify trade-offs. One solution, \ac{DTX} alignment with memory, addresses trade-offs of the other three by maintaining memory of past alignment and channel quality to buffer short term changes in channel quality. All strategies are found to exhibit similar convergence behavior, but different power consumption and retransmission probabilities. \ac{DTX} alignment with memory is shown to achieve up to 40\% savings in power consumption and more than 20\% lower retransmission probability than the \ac{SOTA}.
\end{abstract}

\acresetall


\section{Introduction}
To reduce their power consumption, future \acp{BS} will use very short sleep modes---called \ac{DTX}---which interrupt their transmission~\cite{fmmjg1101}. \ac{DTX} can be employed for energy saving when a communication system has spare capacity, \eg~at night in cellular networks. It can be easily integrated into \ac{OFDMA} systems without incurring delays by not scheduling resources for transmission during one or more time slots and then setting the transmitter to \ac{DTX} mode instead for the duration of these unscheduled time slots. For the transmitter, this is a local power saving measure. When seen from a network perspective, these regular interruptions in transmission can be aligned to reduce interference and power consumption. Such alignment has not been studied in detail, but is related to the field of channel allocation. 


Generally, in channel allocation, the challenge is to assign frequency channels to different cells or links of a network such that mutual interference is minimized. To be applicable for future small-cell cellular networks, such channel allocation needs to be distributed, uncoordinated and dynamic. Distributed operation prevents the delay and backhaul requirements imposed by a central controller. If the allocation is also uncoordinated, it does not require message exchange through a backhaul of limited capacity. Also, as networks, channels and traffic change rapidly, the allocation must be highly dynamic and update often. Problems of this type are known to be NP-hard~\cite{knun01}. Therefore, different suboptimal approaches have been proposed. The simplest approach to distributing channels over a network is a fixed assignment in which it is predefined which links will use which channels, \eg~see~\cite{book:p0001}. However, this is inflexible to changing or asymmetric traffic loads. For flexibility, dynamic channel allocation methods are required. A simple dynamic channel allocation method, Sequential Channel Search~\cite{sg9301}, assigns channels with sufficient quality in a predefined order. This technique allows adjusting the number of channels flexibly, but is suboptimal due to its strong channel overlap between neighbors. Taking the channel quality into consideration by measurement is proposed in the Minimum \ac{SINR} method~\cite{sg9301}. This technique offers increased spectral efficiency, but can lead to instabilities when allocations happen synchronously. Ellenbeck~\ea~\cite{ehb0801} introduce methods of game theory and respond to the instabilities by adding p-persistence to avoid simultaneous bad player decisions. However, the proposed method only applies to single user systems and does not address target rates. Dynamic Channel Segregation~\cite{aa9301} introduces the notion of memory of channel availability, allowing a transmitter to track which channels tend to be favorable for transmission. However, the algorithm only applies to sequential channel decisions based on an idle or busy state in circuit switched networks and cannot be applied to the concurrent alignment of channels as is required in \ac{OFDMA} networks.

In this paper, we combine the findings from past channel allocation research and adapt them for aligning \ac{DTX} time slots in an \ac{OFDMA} frame. We derive four distributed uncoordinated dynamic time slot alignment strategies for the cellular downlink, in which \acp{BS} independently prioritize time slots for transmission while maintaining system stability. The remainder of the paper is structured as follows. Section~\ref{problem} formulates the system model and the problem at hand. The considered four alternative solutions are described in Section~\ref{solutions}. Findings obtained from simulation are presented in Section~\ref{results}. The paper is concluded in Section~\ref{conclusion}.

\section{System model and problem formulation}
\label{problem}

We make the following assumptions about the network. \acp{BS} in a reuse-one \ac{OFDMA} cellular network schedule a target rate, $B_k$, per \ac{OFDMA} frame to be transmitted to each mobile $k=\{1,\dots,K\}$. All time slots are available for scheduling in all cells. Mobiles report perceived \ac{SINR} from the previous \ac{OFDMA} frame, $s_{n,t,k}$, on subcarrier $n=\{1,\dots,N\}$, time slot $t=\{1,\dots,T\}$ to their associated \ac{BS}. 
\acp{BS} have a \ac{DTX} mode available for each time slot during which transmission and reception are disabled and power consumption is significantly reduced to $P_{\mathrm{S}}$ compared to power consumption during transmission, which is a function of the power allocated to each \ac{RB}, $\rho_{\mathrm{Tx}}$. \ac{DTX} is available fast enough to enable it within individual time slots of an \ac{OFDMA} transmission such that transmission time slots are not required to be consecutive. A \ac{BS} can schedule no transmission in one or more time slots and go to \ac{DTX} mode instead. Scheduling a \ac{DTX} time slot in one \ac{BS} reduces the interference on that time slot to all other \acp{BS}. The \ac{OFDMA} frames of all \acp{BS} are assumed to be aligned such that the interference over one time slot and subcarrier is flat and that all \acp{BS} can perform alignment operations in synchrony. The \ac{OFDMA} frame consists of $N T$ \acp{RB}. The channel is subject to block fading. Interference is treated as noise.

To describe the resource allocation problem formally, we define the function 
\begin{equation}
\begin{aligned}
\Pi: \{1,\dots,T\}\times\{1,\dots,N\} &\rightarrow \{0,1,\dots,K\}\\ 
(n,t)&\mapsto k,
\end{aligned}
\end{equation}
which maps to each \ac{RB} a user $k$ or 0, where 0 indicates that the \ac{RB} is not scheduled.

Each cell tries to minimize its power consumption by scheduling time slots for transmission and sleep. With $r_{n,t,k}$ the capacity of resource $(n,t)$ if it is scheduled to $k$, the optimization problem for the total \ac{BS} power consumption, $P_{\mathrm{total}}$, is as follows.
\begin{subequations}
\begin{align}
 \underset{\Pi}{\text{minimize}} \quad P_{\mathrm{total}} &= P_{\mathrm{S}} \frac{T_{\mathrm{S}}}{T} + \rho_{\mathrm{Tx}} N_{\mathrm{Tx}} +P_0\left(\frac{T-T_{\mathrm{S}}}{T}\right) \label{eq:ptotal}&\\
\text{subject to}&&\notag\\ 
\quad N_{\mathrm{Tx}} &= &\label{eq:nt}\\
 |\{ n \in \{1,\dots,N\} &| \left( \Pi(n,t) \neq 0 \quad\forall t \in \{1,\dots,T\}\right)\}|&\notag\\
\quad T_{\mathrm{S}} &= &\label{eq:ts}\\
 |\{ t \in \{1,\dots,T\} &| \left( \Pi(n,t) = 0 \quad\forall n \in \{1,\dots,N\}\right)\}|&\notag\\
  B_k &\leq \displaystyle \sum_{\{(n,t)|\Pi(n,t)=k\}} r_{n,t,k} \quad \forall k.&\label{eq:bk}
\end{align}
\end{subequations}
With $|\cdot|$ the cardinality operator, $T_{\mathrm{S}}$ is the number of time slots in which for all $n$, no user is allocated and the \ac{BS} can enable \ac{DTX} and $N_{\mathrm{Tx}}$ is the number of \acp{RB} that are scheduled for transmission. The constraint~\eqref{eq:bk} provides the rate guarantee for each user.

The difficulty lies in finding the mapping $\Pi$. 
Under a brute force approach, there exist $(K+1)^{NT}$ possible combinations. Due to the large number \ac{RB}s present in typical \ac{OFDMA} systems like \ac{LTE}, the computation of the solution is infeasible. Consequently, we resort to heuristic methods in the next section.

Some of the methods compared in this paper make use of a time slot ranking. In other works, \eg~\cite{aa9301}, ranking is proposed to be based on \ac{SINR}. However, in a multi-user \ac{OFDMA} system each subcarrier and mobile terminal have a different \ac{SINR}, thus generating a problem of comparability between time slots. Consequently, we propose to compare time slots by their hypothetical sum capacity, $B_t$, over all mobiles and subcarriers with
\begin{equation}
\label{eq:bt}
\begin{aligned}
B_{t} &= \sum_{k=1}^K \sum_{n=1}^N \log_2 \left( 1 + s_{n,t,k} \right).
\end{aligned}
\end{equation}

\section{\ac{DTX} alignment strategies}
\label{solutions}

In this section, we derive four strategies to tackle the problem at hand which differ in performance and complexity.

\subsection{Sequential alignment}
This strategy is to always allocate as many time slots for transmission as required in sequential order and set the remainder to \ac{DTX}. This leads to strongest overlap and consequently highest interference on the first time slot and lowest overlap and possibly no transmissions at all on the last time slot. This strategy does not make use of the available channel quality information and provides valuable insights into questions of stability, reliability and convergence. It serves as a deterministic upper bound.

\subsection{Random alignment}
Random alignment refers to randomly selecting transmission time slots for every \ac{OFDMA} frame and setting the remainder to \ac{DTX}. This strategy provides a reference for achievable gains, allows to assess the worst cast effects of randomness and represents the state of the art in today's unsynchronized unaligned networks.

\subsection{P-persistent ranking}
The synchronous alignment of uncoordinated \acp{BS} can lead to instabilities, when neighboring \acp{BS}--perceiving similar interference information--schedule the same time slots for transmission. This leads to oscillating scheduling which never reaches the desired system state~\cite{ehb0801}. To address this problem, we introduce p-persistence to break the unwanted synchrony by only changing established \ac{DTX} schedules with a probability $p=0.3$. In initial tests, the value of $p$ did not have a significant effect on the power consumption. The exact tuning of $p$ is out of the scope this paper.  P-persistent ranking first ranks time slots by their sum capacity as in~\eqref{eq:bt}, schedules time slots in that order, but only applies this new selection with probability $p$. Otherwise, the schedule from the last iteration remains active. After ranking, time slots are selected for transmission in order of $B_t$. The remainder of time slots is set to \ac{DTX}.

\subsection{Distributed \ac{DTX} alignment with memory}
To counter oscillation and achieve a convergent network state we introduce memory of past schedules into the alignment process. The rationale behind the distributed \ac{DTX} alignment with memory algorithm is as follows. Taking into account current time slot capacities, $B_t$, past scores and slot allocations, each \ac{BS} first updates the internal score of each time slot and then returns the priority of time slots by score. In case of equal scores, time slots are further sub-sorted by $B_t$. The score is updated in integers. All time slots which were used for transmission in the previous \ac{OFDMA} frame receive a score increment of one. Furthermore, the time slot with highest $B_t$, $R_0$, receives an increment of one. Time slots which were not used for transmission in the previous \ac{OFDMA} frame receive a decrement of one, except for $R_0$. Scores have an upper limit $\psi_{\mathrm{ul}}$ beyond which there is no increment and a lower limit $\psi_{\mathrm{ll}}$ below which there is no decrement. The difference between $\psi_{\mathrm{ul}}$ and $\psi_{\mathrm{ll}}$ can be interpreted as the depth of the memory buffer.

\begin{algorithm}
\caption{Distributed DTX alignment with memory}
\label{alg1}
\begin{algorithmic}[1]
\ENSURE {$\psi, \Upsilon_{\mathrm{u}}$}
\STATE $R \leftarrow $sort-desc-by-capacity$(\Upsilon)$
\FORALL{$\upsilon$ in $\Upsilon_{\mathrm{u}}$}
  \IF {$\psi (\upsilon) < \psi_{\mathrm{ul}}$}
    \STATE $\psi(\upsilon) \leftarrow \psi(\upsilon) + 1$
  \ENDIF
\ENDFOR
\FORALL{$\upsilon$ in $\Upsilon_{\mathrm{uu}} \backslash \{R_0\}$}
  \IF {$\psi (\upsilon) > \psi_{\mathrm{ll}}$}
    \STATE $\psi(\upsilon) \leftarrow \psi(\upsilon) - 1$
  \ENDIF
\ENDFOR
\STATE $\psi (R_0) \leftarrow \psi(R_0) + 1$
\STATE $V \leftarrow $ sort-desc-by-score$(R, \psi, \Upsilon)$
\RETURN $\psi, V, \Upsilon_{\mathrm{u}}$
\end{algorithmic}
\end{algorithm}

The described steps are summarized in Algorithm~\ref{alg1} with scoring map $\psi$, ranking tuple $R$, and priority tuple $V$. The set of used time slots $\Upsilon_{\mathrm{u}}$ and unused time slots $\Upsilon_{\mathrm{uu}}$ make up the set of time slots $\Upsilon$, the cardinality of which is $T$. The function 'sort-desc-by-capacity($\Upsilon$)' returns a list of time slots ordered descending by $B_t$. The function 'sort-desc-by-score$(R, \psi, \Upsilon)$' returns a list of time slots ordered primarily by descending score and secondarily by descending $B_t$. After the execution of Algorithm~\ref{alg1}, the first $T_{\mathrm{Tx}}$ time slots in $V$ are scheduled for transmission.

In the following, we illustrate the iterations over three \ac{OFDMA} frames of Algorithm~\ref{alg1} for a system with three time slots. In the example, the algorithm delays the change of $\Upsilon_{\mathrm{u}}$ from $c$ to $b$ to buffer scheduling changes. The iteration begins with arbitrarily chosen $\Upsilon=\{a,b,c\}$, $\psi=\{a:0,b:2,c:5\}$:
\begin{enumerate}
 \item $\Upsilon_{\mathrm{u}}=\{c\}$, $R = ( b,c,a ) $\\
$\rightarrow \psi = \{a:0,b:3,c:5\}$, $V = (c,b,a)$
 \item $\Upsilon_{\mathrm{u}}=\{b,c\}$, $R = ( b,c,a ) $\\
$\rightarrow \psi = \{a:0,b:5,c:5\}$, $V = (b,c,a)$
 \item $\Upsilon_{\mathrm{u}}=\{b\}$, $R = ( b,a,c ) $\\
$\rightarrow \psi = \{a:0,b:5,c:4\}$, $V = (b,c,a)$
\end{enumerate}

When applied, Algorithm~\ref{alg1} strongly benefits time slots which were used for transmission in the past ($b,c$ in step 1 of the example). These time slots tend to repeatedly receive score increments until they all have maximum score $\psi_{\mathrm{ul}}$ ($b,c$ in step 2 of the example). When the score is equal for some time slots ($b,c$ in step 2 of the example), the ranking is based on $B_t$. A time slot which was used for transmission and has highest  $B_t$, $R_0$, receives the highest increment (slot $b$ in step 2 of the example). When a time slot has highest  $B_t$, but was not selected for transmission in the previous \ac{OFDMA} frame, it receives a score increment, but is not guaranteed to be used for transmission (slot $b$ in step 1 of the example). When a time slot repeatedly has highest $B_t$, it reaches $\psi_{\mathrm{ul}}$ ($b$ in the example). The algorithm thus buffers short term changes in the channel quality setting in favor of long term time slot selection.

\section{Results}
\label{results}
We analyze the four alignment strategies using computer simulations with regard to power consumption, convergence, reliability of delivered rates and algorithmic complexity after the introduction of the simulation environment and the \ac{RB} scheduling scheme.

\subsection{Simulation environment and \ac{RB} scheduling}
The four strategies were tested in a network simulation with 19-cell hexagonal arrangement with uniformly distributed mobiles and fixed target rates per mobile. Data was collected only from the center cell which is thus surrounded by two tiers of interfering cells. Power consumption of a cell is modeled as a function of transmission power as described in~\cite{ddgfahwsr1201}. Table~\ref{tab:params} lists additional parameters used which approximate an \ac{LTE} system. The simulation is started with the assumption of full transmission power on all resources with a power consumption of 350~W per cell as a worst-case initial configuration.

\small
\begin{table}
\centering
      \begin{tabular}{c|c}
	Parameter          			& Value\\
	\hline
	Carrier frequency  			& 2\,GHz\\
	Intersite distance 			& 500\,m\\
	Pathloss model  			& 3GPP UMa, NLOS, \\
						& shadowing~\cite{std:3gpp-faeutrapla}\\
	Shadowing standard deviation		& 8\,dB\\
	Bandwidth				& 10\,MHz \\
	Transmission power per \ac{RB}	& 0.8~W \\
	Thermal noise temperature		& 290\,K \\
	Interference tiers			& 2 (19 cells)\\
	Mobile target rate			& 2~Mbps and 3~Mbps\\
	\ac{OFDMA} subframes (time slots)	& 10\\
	Subcarriers				& 50\\
	Mobiles					& 10\\
	$\psi_{\mathrm{ul}}$			& 5\\
	$\psi_{\mathrm{ll}}$			& 0\\
	Power model factors~\cite{ddgfahwsr1201}\\ (idle; load factor; DTX)& 200~W; 3.75; 90~W
	\end{tabular}
\label{tab:params}
      \caption{Simulation Parameters}
\end{table}
\normalsize
In order to assess the performance of the four strategies, it is necessary to make assumptions about how the individual \ac{RB}s are scheduled within a time slot. In order to avoid masking effects of the algorithms under test, we have applied a sequential \ac{RB} allocation. Sequential resource allocation is performed after the \ac{DTX} alignment step is completed and allocates as many bits to an \ac{OFDMA} resource block as possible according to the Shannon capacity, followed in order by the next resource in the same time slot (sequentially), until the target rate has been scheduled to each mobile. Time slots are scheduled in the order provided by each of the four strategies. This sequential resource scheduler deliberately omits the benefits of multi-user diversity. This leads to underestimating achievable rates in simulation compared to a system which exploits multi-user diversity, but allows a fair comparison of the quality of different \ac{DTX} alignment algorithms.

\subsection{Power consumption}
\begin{figure}
\centering
\includegraphics[width=0.4\textwidth]{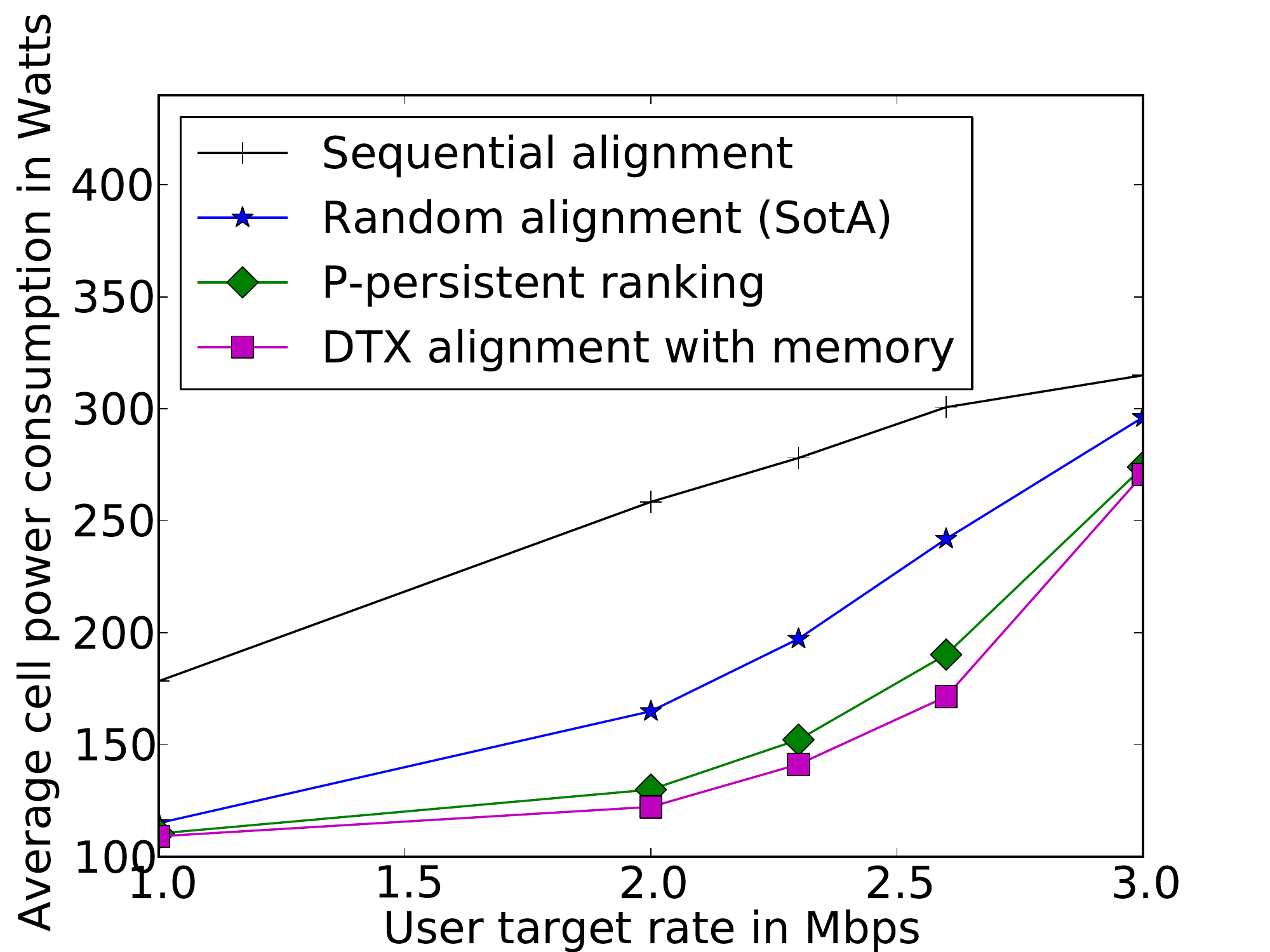}
 \caption{\ac{BS} power consumption over different cell sum rates.}
\label{fig:sumrate}
\end{figure}

To assess achievable power savings and the dynamic adaptivity over a large range of cell loads, we inspect the cell total power consumption in Fig.~\ref{fig:sumrate}. 
At low load very few \ac{RB}s are required to deliver the target rate and more time slots can be scheduled for \ac{DTX} than at high traffic loads, leading to monotonously rising power consumption over increasing target rates for all alignment methods. 

Sequential alignment causes the highest power consumption over any target rate with an almost linear relationship between user target rates and power consumption. Sequential alignment consumes high power, as it schedules many \ac{RB}s to achieve the target rate due to the high interference level present. 

This power consumption is significantly lower for the \ac{SOTA}, random alignment. The randomness of time slot alignment creates a much lower average interference than with sequential alignment allowing more data to be transmitted in each \ac{RB}. As fewer \ac{RB}s are required, less power is consumed.

P-persistent ranking and \ac{DTX} with memory both achieve similarly low power consumption of up to 40\% less than random alignment. The relationship between power consumption and target rate is noticeably non-linear, as it is flat at low target rates and grows more steeply at high target rates. This behavior is caused by the low interference level these strategies manage to create. Only at high rates, when the number of sleep time slots has to be significantly decreased, does the interference increase, leading to higher power consumption.

Also noteworthy is the fact that at 1~Mbps and 3~Mbps, random alignment performs nearly as good as p-persistent ranking. At these extreme points the network is almost unloaded and almost fully loaded, respectively. Consequently, either most time slots are scheduled for \ac{DTX} or none, leaving very little room for optimization compared to randomness. The largest potential for time slot alignment for power saving is in networks which are medium loaded. Under medium load, the number of transmission and \ac{DTX} time slots is similar, causing the effects of alignment to be most pronounced.

\subsection{Convergence}
\begin{figure}
\centering
\includegraphics[width=0.4\textwidth]{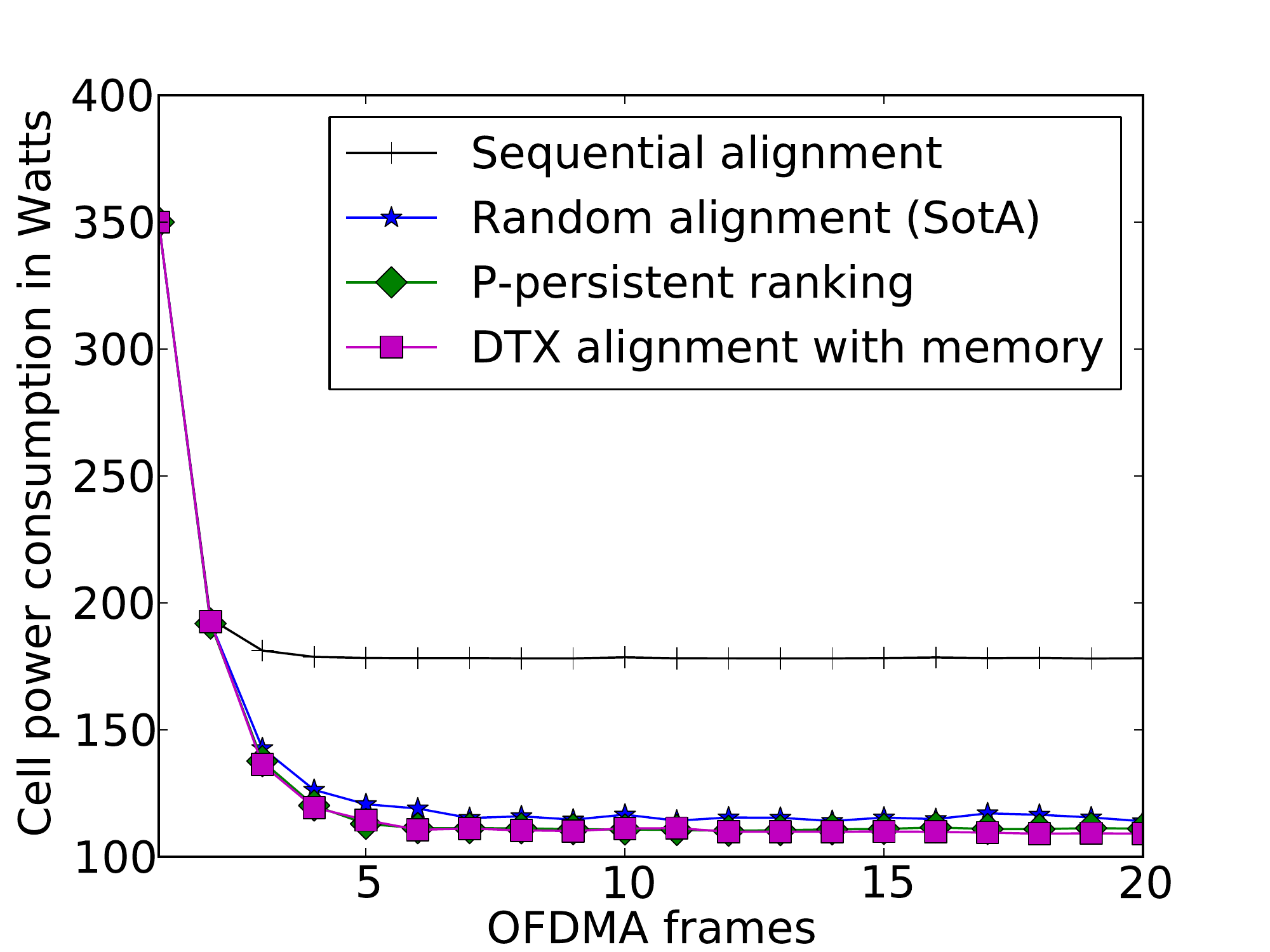}
 \caption{\ac{BS} power consumption over \ac{OFDMA} frames at 1~Mbps per mobile.}
\label{fig:power_consumption_1000000}
\end{figure}

\begin{figure}
\centering
\includegraphics[width=0.4\textwidth]{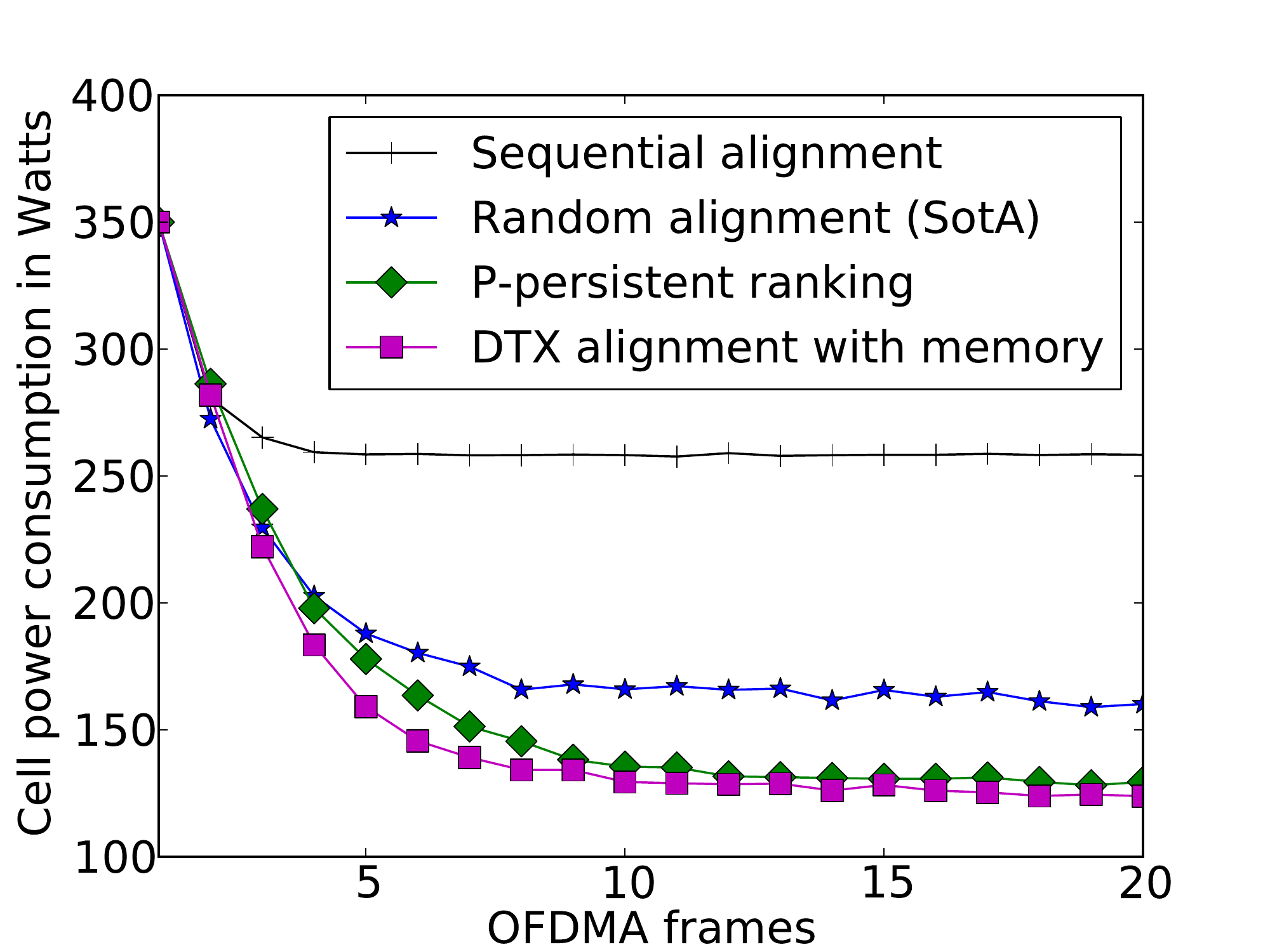}
 \caption{\ac{BS} power consumption over \ac{OFDMA} frames at 2~Mbps per mobile.}
\label{fig:power_consumption_2000000}
\end{figure}

Another relevant aspect is the convergence of the network to a stable state. As each \ac{BS} makes iterative adjustments to its selection of time slots for transmission, the speed of convergence as well as the convergence to a stable point of operation are relevant criteria. The effect of the iterative execution of the four strategies on the cell power consumption is illustrated in Fig.~\ref{fig:power_consumption_1000000}. Power consumption is found to converge to a stable value within within six \ac{OFDMA} frames (alignment iterations). All strategies converge to the average power consumption values shown in Fig.~\ref{fig:sumrate}. The simulation starts from a worst-case schedule of transmission on all \ac{RB}s and then iteratively schedules time slots for transmission and \ac{DTX}. In the case of 1~Mbps per user, one transmission time slot is sufficient to schedule the target rate. With transmissions only taking place during one time slot, there is very little difference between a random alignment and p-persistent ranking or \ac{DTX} with memory. At 2~Mbps per user, see Fig.~\ref{fig:power_consumption_2000000}, random alignment occasionally causes higher interference than p-persistent ranking or \ac{DTX} with memory, leading to higher power consumption. Also, p-persistent ranking converges more slowly than \ac{DTX} alignment with memory. 

\subsection{Reliability}
\begin{figure}
\centering
\includegraphics[width=0.4\textwidth]{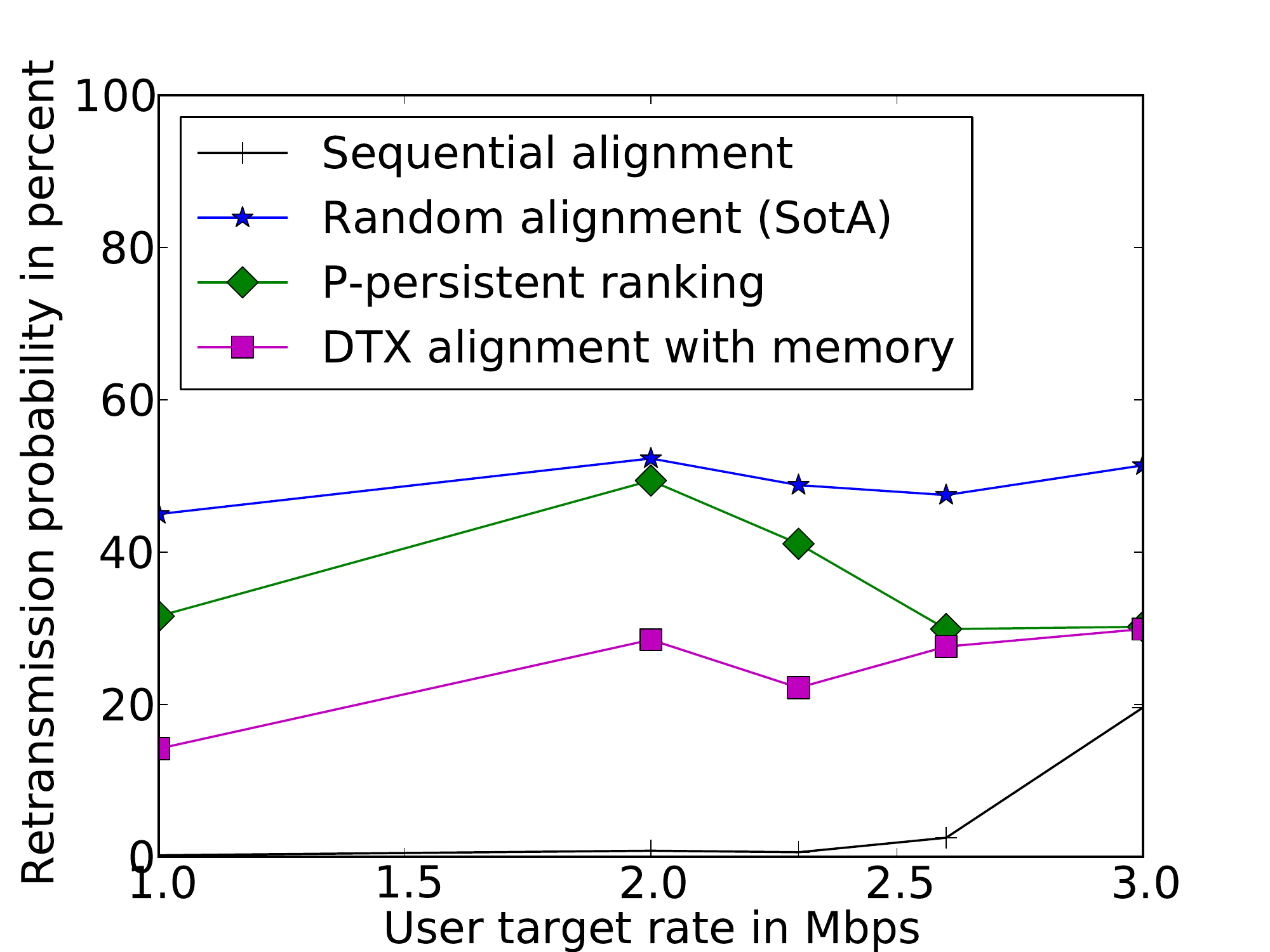}
 \caption{Retransmission probability over targeted rate.}
\label{fig:percentage_satisfied_over_target_rate}
\end{figure}

An important aspect in dynamical systems with target rates is that scheduled target rates cannot always be fulfilled. As \ac{RB} scheduling is based on channel quality information which was collected in the previous \ac{OFDMA} frame, the actual channel quality during transmission may differ, leading to lower than expected rates. Thus, although certain target rates are scheduled and although the system is not fully loaded, a \ac{BS} may fail to deliver the targeted rate and require retransmission of some \acp{RB}. We assess this metric by considering the retransmission probability for each strategy. The results are shown in Fig.~\ref{fig:percentage_satisfied_over_target_rate} against a range of user target rates. 

Easiest to interpret is sequential alignment which does not require retransmission for target rates up to 2.3~Mbps due to its determinism. The increase in the retransmission probability at high rates is not caused by a failure of the alignment, but by system overload. When high rates are combined with bad channel conditions, the system may be unable to deliver the target data rate, independent of the alignment strategy. This increase of the retransmission probability at high rates is present for all alignment strategies and constitutes outage.

The retransmission probability is highest for random alignment. This is caused by the strong fluctuation of interference under randomized scheduling. Channel quality measurements used for \ac{RB} scheduling are of very little reliability, as the interference changes quickly due to random time slot alignment, resulting in increased retransmission probability.

P-persistent ranking performs slightly better than random alignment at 1~Mbps target rate. At 2~Mbps per user, where the alignment potential is highest, oscillation causes the highest retransmission probability.

\ac{DTX} alignment with memory achieves a much lower retransmission probability in the range of 15\% to 20\%, due to the reduction in interference fluctuation introduced by the memory score. 

\subsection{Complexity}
With regard to complexity, sequential and random alignment are of minimal complexity. These strategies involve no algorithmic decision-making on the set of transmission time slots. P-persistent ranking requires one ranking and time slot selection with probability $p$ per iteration. The highest complexity is present in \ac{DTX} with memory, which requires two executions of the time slot sort, one for the generation of the score and one for the output of the ranking. Although \ac{DTX} with memory comprises the highest complexity in comparison, these operations pose a small burden on modern hardware as the number of time slots is typically small. For example, typical \ac{LTE} systems are designed with 10~subframes.

\subsection{Interpretation}
To conclude our evaluation, we have found that \ac{DTX} with memory provides the best results. Under the present assumptions, it provides both a lower power consumption and lower retransmission probability than the \ac{SOTA} and p-persistent ranking. Future \ac{DTX} capable networks can and should exploit this alignment potential.

\section{Conclusion}
\label{conclusion}
In this paper, we have studied the constructive uncoordinated and distributed alignment of \ac{DTX} time slots between interfering \ac{OFDMA} \acp{BS} for power-saving under rate constraints. We first established the open problem formally. Due to its complexity, we have approached it with four alternative heuristic strategies. One of these strategies, \ac{DTX} with memory, is an original contribution of this paper and introduces memory to overcome short-term fluctuations and networks oscillations. The performance of the four strategies was tested in simulation. It was found that power consumption can be reduced significantly, especially at medium cell loads. All strategies converge within six \ac{OFDMA} frames. \ac{DTX} with memory reduces power consumption up to 40\% compared to the \ac{SOTA}, combined with around 20\% reduction in retransmission probability.



\bibliography{2014icc} 
\bibliographystyle{ieeetran}

\end{document}